\documentclass{appolb}
\usepackage{graphicx}
\usepackage{hyperref}
\usepackage{xcolor}
\definecolor{linkcolor}{HTML}{8b0000} 
\definecolor{urlcolor}{HTML}{8b0000} 
\definecolor{citecolor}{HTML}{8b0000}
\hypersetup{pdfstartview=FitH, linkcolor=linkcolor, urlcolor=urlcolor, citecolor=citecolor, colorlinks=true}

\usepackage[
    backend=bibtex, 
    articletitle=false,
    style=phys,
    biblabel=brackets,
    doi=false, url=false, isbn=false,
    hyperref=auto
]{biblatex}

\begin{document}

\title{\uppercase{Studies of the linearity and stability of Silicon Drift Detectors for kaonic atoms X-ray spectroscopy}%
\thanks{Presented at the 4th Jagiellonian Symposium on Advances in Particle Physics and Medicine, Kraków, Poland, 10-15 July 2022.}%
}

\author{A.~Khreptak$^{a,b,}$\thanks{Corresponding author: \texttt{aleksander.khreptak@lnf.infn.it}}, C.~Amsler$^c$, M.~Bazzi$^a$, D.~Bosnar$^d$, M.~Bragadireanu$^e$, M.~Carminati$^{f,g}$, M.~Cargnelli$^c$, A.~Clozza$^a$, G.~Deda$^{f,g}$, L.~De~Paolis$^a$, R.~Del~Grande$^{h,a}$, L.~Fabbietti$^h$, C.~Fiorini$^{f,g}$, C.~Guaraldo$^a$, M.~Iliescu$^a$, M.~Iwasaki$^i$, S.~Manti$^a$, J.~Marton$^c$, M.~Miliucci$^a$, P.~Moskal$^{b,j}$, F.~Napolitano$^a$, S.~Nied\'{z}wiecki$^{b,j}$, H.~Ohnishi$^k$, K.~Piscicchia$^{l,a}$, Y.~Sada$^k$, A.~Scordo$^a$, F.~Sgaramella$^a$, H.~Shi$^c$, M.~Silarski$^{b,j}$, D.~Sirghi$^{a,l,e}$, F.~Sirghi$^{a,e}$, M.~Skurzok$^{b,j}$,  A.~Spallone$^a$, K.~Toho$^k$, M.~T{\"u}chler$^{c,m}$, O.~Vazquez~Doce$^a$, J.~Zmeskal$^c$, C.~Yoshida$^k$, and~C.~Curceanu$^a$ 
\address{$^a$ INFN, Laboratori Nazionali di Frascati, Frascati (Roma), Italy \\
$^b$ Institute of Physics, Jagiellonian University, Krak\'{o}w, Poland \\
$^c$ Stefan-Meyer-Institut f\"{u}r Subatomare Physik, Vienna, Austria \\
$^d$ Department of Physics, Faculty of Science, University of Zagreb, Croatia \\
$^e$ IFIN-HH, Institutul National pentru Fizica si Inginerie Nucleara Horia Hulubbei, M\u{a}gurele, Romania \\
$^f$ Politecnico di Milano, Dipartimento di Elettronica, Informazione e Bioingegneria, Milano, Italy \\
$^g$ INFN Sezione di Milano, Italy  \\
$^h$ Physik Department E62, Technische Universit\"{a}t M\"{u}nchen, Garching, Germany \\
$^i$ RIKEN, Institute of Physical and Chemical Research, Wako, Japan \\
$^j$ Center for Theranostics, Jagiellonian University, Krak\'{o}w, Poland \\
$^k$ Research Center for Electron Photon Science (ELPH), Tohoku University, Sendai, Japan \\
$^l$ Centro Ricerche Enrico Fermi - Museo Storico della Fisica e Centro Studi e Ricerche ``Enrico Fermi", Rome, Italy \\
$^m$ University of Vienna, Vienna Doctoral School in Physics, Vienna, Austria } 
}


\maketitle

\begin{abstract}
The SIDDHARTA-2 experiment at the DA$\Phi$NE collider aims to perform precision measurements of kaonic atoms X-ray spectroscopy  for the investigation of the antikaon-nucleon strong interaction. 
To achieve this goal, novel large-area Silicon Drift Detectors (SDDs) have been developed.
These devices have special geometry, field configuration and readout electronics that  ensure excellent performance in terms of linearity and stability.
The paper presents preliminary results for the linearity determination and stability monitoring of the SDDs system during the measurement of kaonic deuterium carried out in the summer of 2022. 
\end{abstract}

\small{DOI:10.5506/APhysPolBSupp...}
  
\section{Introduction}

The Silicon Drift Detectors (SDDs)~\cite{LECHNER1996346,LECHNER2001281} combine the silicon p–n junction response to ionizing radiation with an innovative electronic field design~\cite{Miliucci_MST2022}.
These devices play a key role in high-precision X-ray spectroscopy of light exotic atoms~\cite{BAZZI2011264}, due to their low electronic noise, good energy resolution and high rate capability~\cite{Miliucci_MST2021}. 
In particular, their excellent capabilities are exploited to extract the shift~($\epsilon$) and width~($\Gamma$) of atomic levels caused by the strong antikaon-nucleon interaction~\cite{Miliucci_CM2021}. 
Experimental studies of light kaonic atoms provide a unique tool to investigate the non-perturbative quantum chromodynamics (QCD) in the strangeness sector~\mbox{\cite{Curceanu_RMP2019,Curceanu_Sym2020}}.

After the successful data taking campaign of the SIDDHARTA (Silicon Drift Detectors for Hadronic Atom Research by Timing Application) experiment in 2009, resulting in the most precise measurement of the kaonic hydrogen~($\overline{K}p$) 
fundamental level shift~($\epsilon$) and width~($\Gamma$)~\cite{BAZZI2011113}, the SIDDHARTA-2 collaboration implemented several updates of the apparatus to perform the analogous, more challenging kaonic deuterium~($\overline{K}d$) 
$2p\rightarrow1s$ transition measurement~\cite{Tuchler_EPJWC2022}.
The experimental facility is located at the DA$\Phi$NE collider~\cite{Milardi_IPAC2018} of the Istituto Nazionale di Fisica Nucleare - Laboratori Nazionali di Frascati (INFN-LNF) in Italy. 

\begin{figure}[htb]
    \centering
    \includegraphics[width=0.45\textwidth]{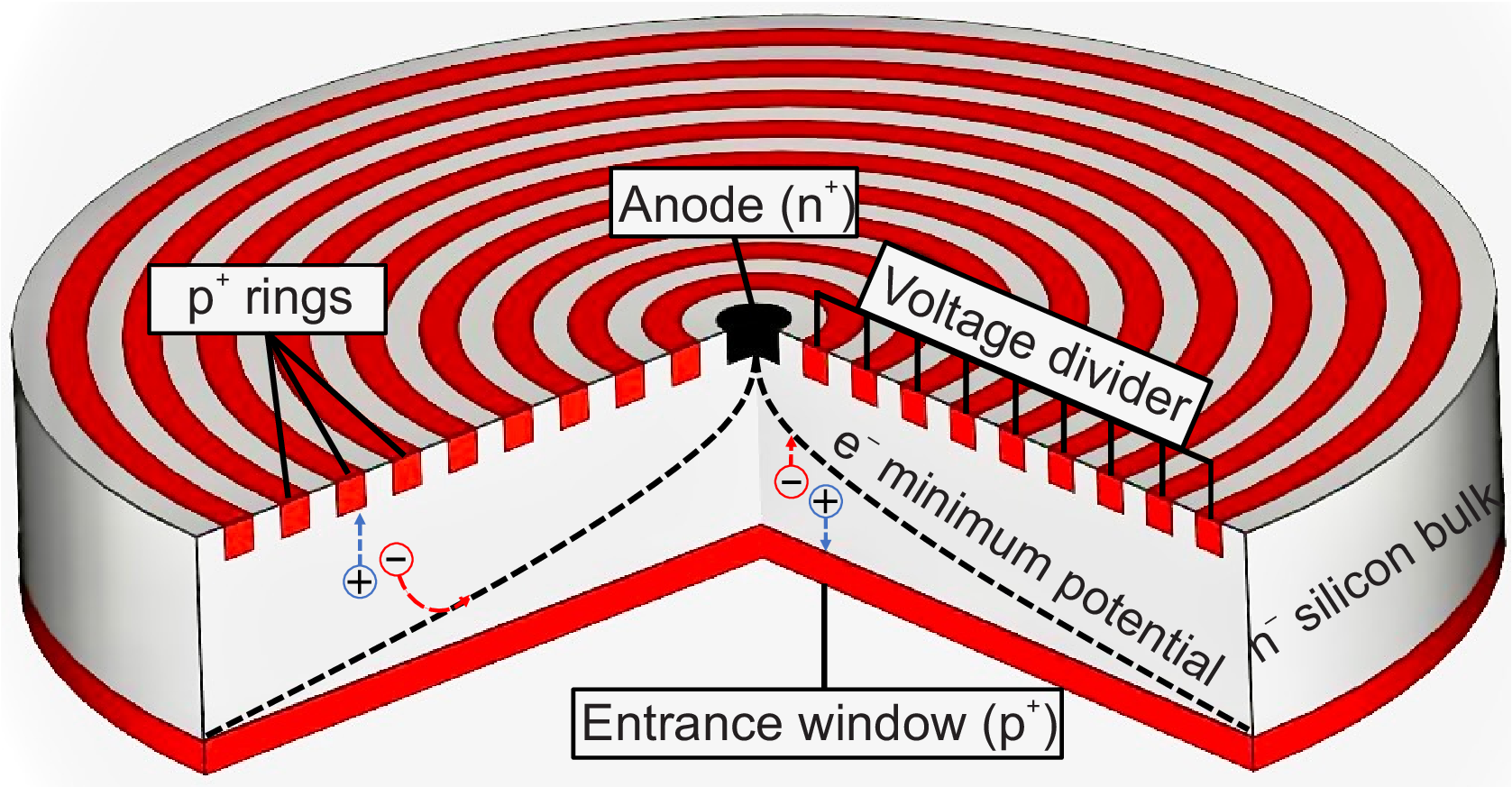}
    \caption{Schematic layout of a cylindrical SDD. The dotted lines represent the drift field inside the bulk. Figure adapted from Ref.~\cite{Sgaramella_arxiv}}
    \label{Fig:SDD}
\end{figure}

The SDD detector, used in the experiment, is composed of a fully depleted n$^{-}$ silicon cylindrical bulk, double sided by p$^{+}$ silicon doped electrodes~\mbox{\cite{LECHNER1996346,LECHNER2001281}}. 
On one side, a layer of p$^{+}$ concentric ring strips are biased with a negative voltage through an internal divider, while on the opposite side, a metalized non-structured p$^{+}$ layer forms the radiation entrance window (see Fig.~\ref{Fig:SDD}).
The electrons produced by the absorbed radiation are focused by the radial drift field to the small n$^{+}$ anode, placed in the center of the structured side and connected to the dedicated front-end electronics. 

New monolithic large-area SDD arrays have been developed by Fondazione Bruno Kessler, together with Politecnico di Milano, INFN-LNF in Italy and the Stefan Meyer Institute (SMI) in Austria, to perform high precision kaonic deuterium spectroscopy~\cite{Miliucci_MST2022,Curceanu_RMP2019}. 
The 450 $\mu$m thick SDD array consists of a $2\times4$ matrix of square cells (0.64 cm$^2$ area for each single unit), mounted on an alumina carrier.
%
The signals coming from the detectors are buffered by a CMOS low-noise charge sensitive preamplifier (CUBE~\cite{CUBE}), directly connected to the n$^{+}$ anode, then processed by a dedicated analog SDD front-end readout ASIC (SFERA~\cite{SFERA}). 

The SIDDHARTA-2 experimental apparatus~\cite{Miliucci_CM2021,Tuchler_EPJWC2022} 
consists of 48 SDD arrays, resulting in a total of 384 readout channels, and a luminometer built with plastic scintillators, based on J-PET technology~\cite{Skurzok_JINST2020,JPET,Moskal_PMB2021,Moskal_Nat2021,Moskal_SciA2021}.
Studies of the linearity and stability of the SDD system are mandatory to achieve the planned high precision measurement of kaonic deuterium X-rays by keeping the systematic error within a~few~eV.


\section{Linearity characterization}
\label{Sec:Linearity}

\begin{figure}[htb]
\centerline{%
\includegraphics[width=0.49\textwidth]{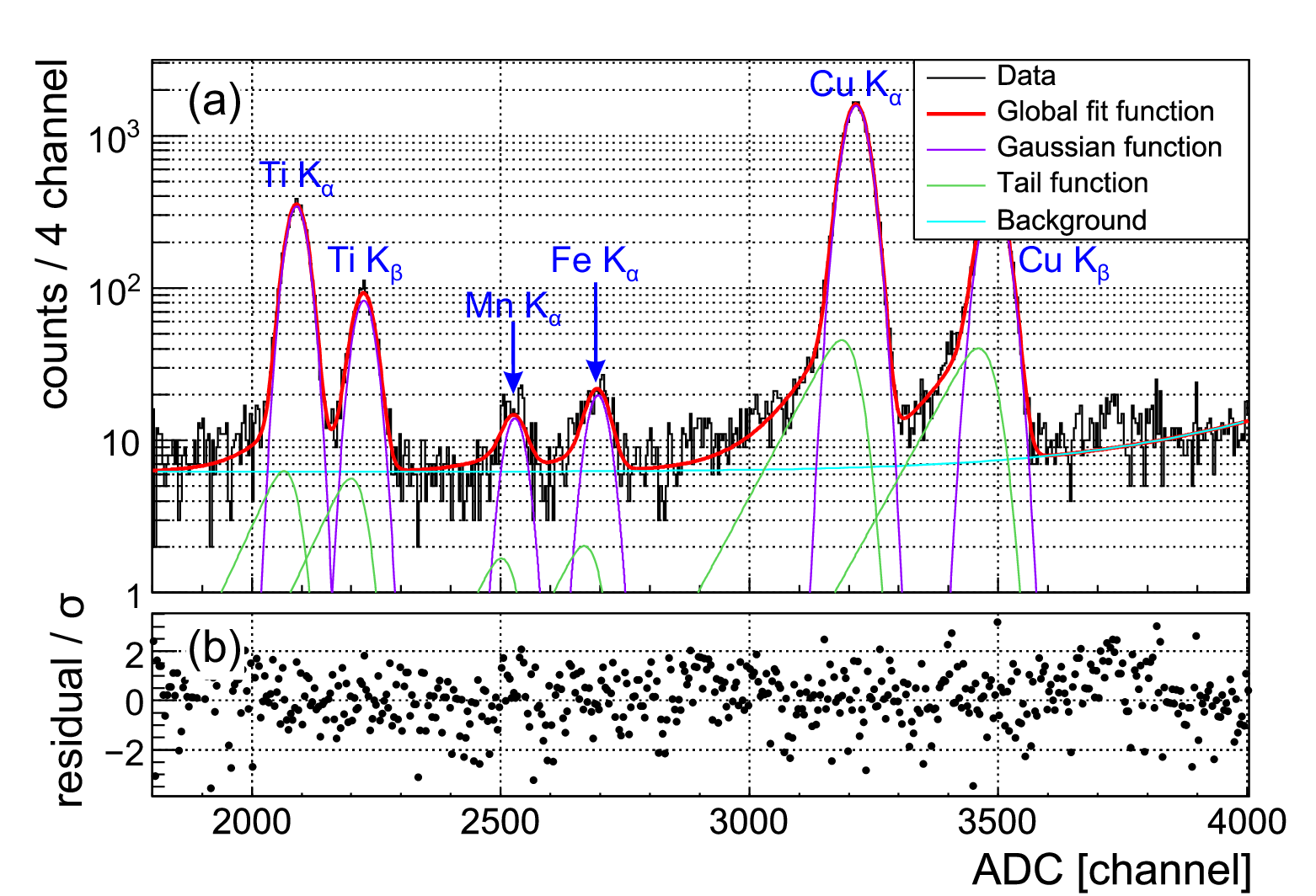}
\includegraphics[width=0.51\textwidth]{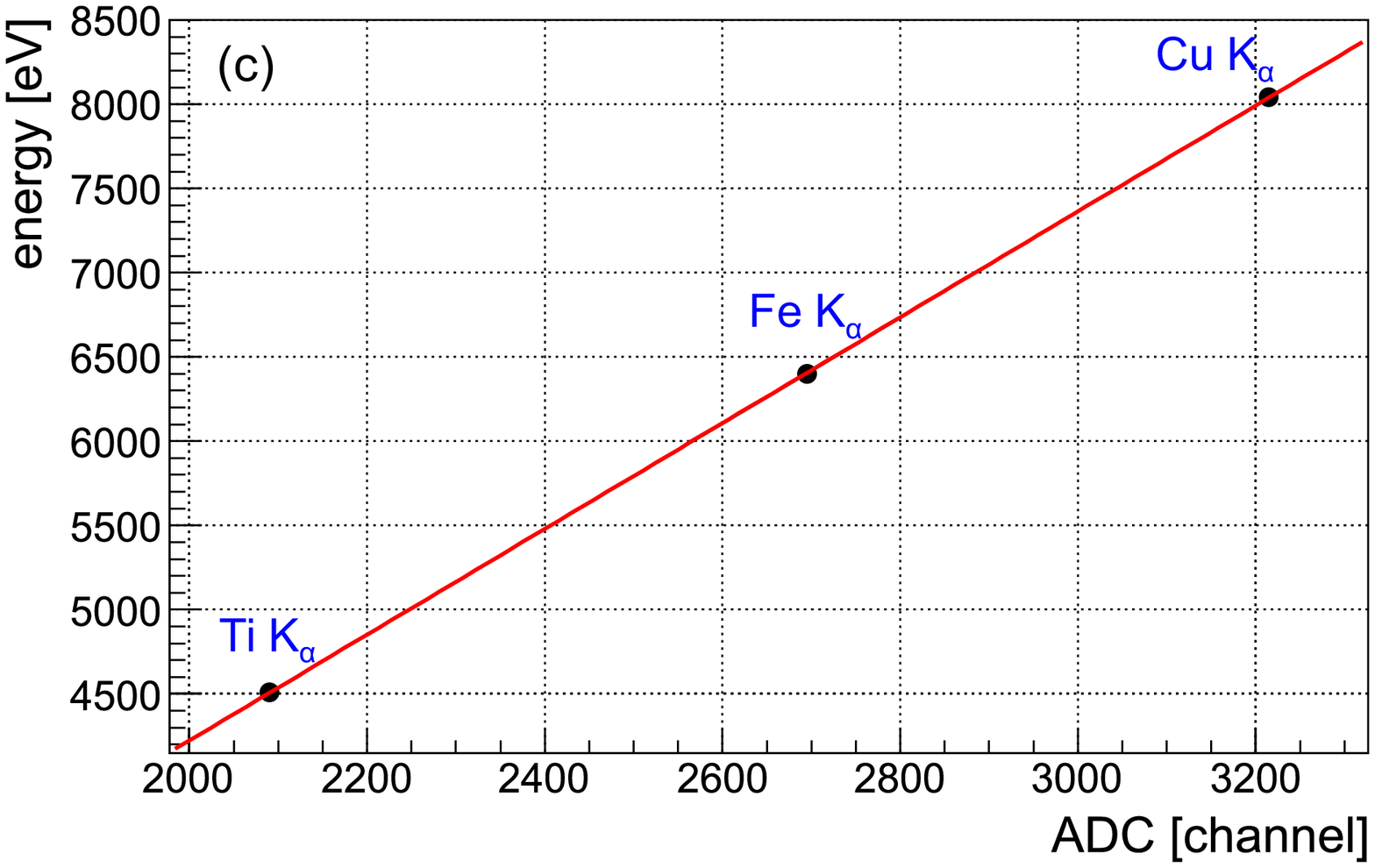}}
\vspace{-2pt}
\caption{(a) A fitted fluorescence spectrum for a single SDD in arbitrary ADC units. 
Red line shows the overall fit function consisting of a sum of Gaussian (violet line) and tail functions (green line) for each peak, together with a constant plus an exponential function to describe the background (cyan line).
(b) Residuals from the fit of the spectrum.
(c) Linear interpolation of the experimental K$_{\alpha}$ peak positions as a function of the known energies of transitions.}
\label{Fig:hADC}
\vspace{-8pt}
\end{figure}

The SDD linearity studies were performed with an X-ray tube exciting a target made of high purity titanium and copper strips~\cite{Sgaramella_arxiv}. 
Fig. \ref{Fig:hADC}(a) shows a typical spectroscopic response of a single SDD unit at $-145^{\circ}$C, irradiated by fluorescence emission from the target strips, activated by the \mbox{X-ray} tube operated at 25 kV and 50 $\mu$A. 
Together with the titanium (Ti) and copper (Cu) lines, the Mn~K$_{\alpha}$ and Fe~K$_{\alpha}$ (produced by the excitation of other components of the setup) are also visible. 

The spectral response of silicon detectors is predominantly described by a Gaussian function for each X-ray peak, with a low energy tail due to the incomplete charge collection at the anode, and electron-hole recombination. 
Thus, the shape of the peaks was modeled taking into account the intrinsic Gaussian statistical spread and the tail effect~\cite{Campbell_NIMPR1990,Campbell_NIMPR1997,Gysel_XRS2003}. 

The fit function (red curve in Fig.~\ref{Fig:hADC}(a)) reproduces the fluorescence lines coming from the target.  
The various contributions to the overall fit function are also shown. 
A constant plus an exponential function describes the background.
Fig.~\ref{Fig:hADC}(b) presents the fit residuals.
The Ti K$_{\alpha}$, Fe K$_{\alpha}$ and Cu K$_{\alpha}$ peaks were exploited to evaluate the linearity of the detectors, due to their higher signal-to-background ratio.
The positions of these peaks were plotted as a function of the tabulated energy values~\cite{xrayBooklet}, and a straight line was used to interpolate the points (see Fig.~\ref{Fig:hADC}(c)). 
The slope of the fit function (in eV/channel) gives the gain parameter value of the spectrum. 

\begin{figure}[htb]
\centerline{%
\includegraphics[width=0.49\textwidth]{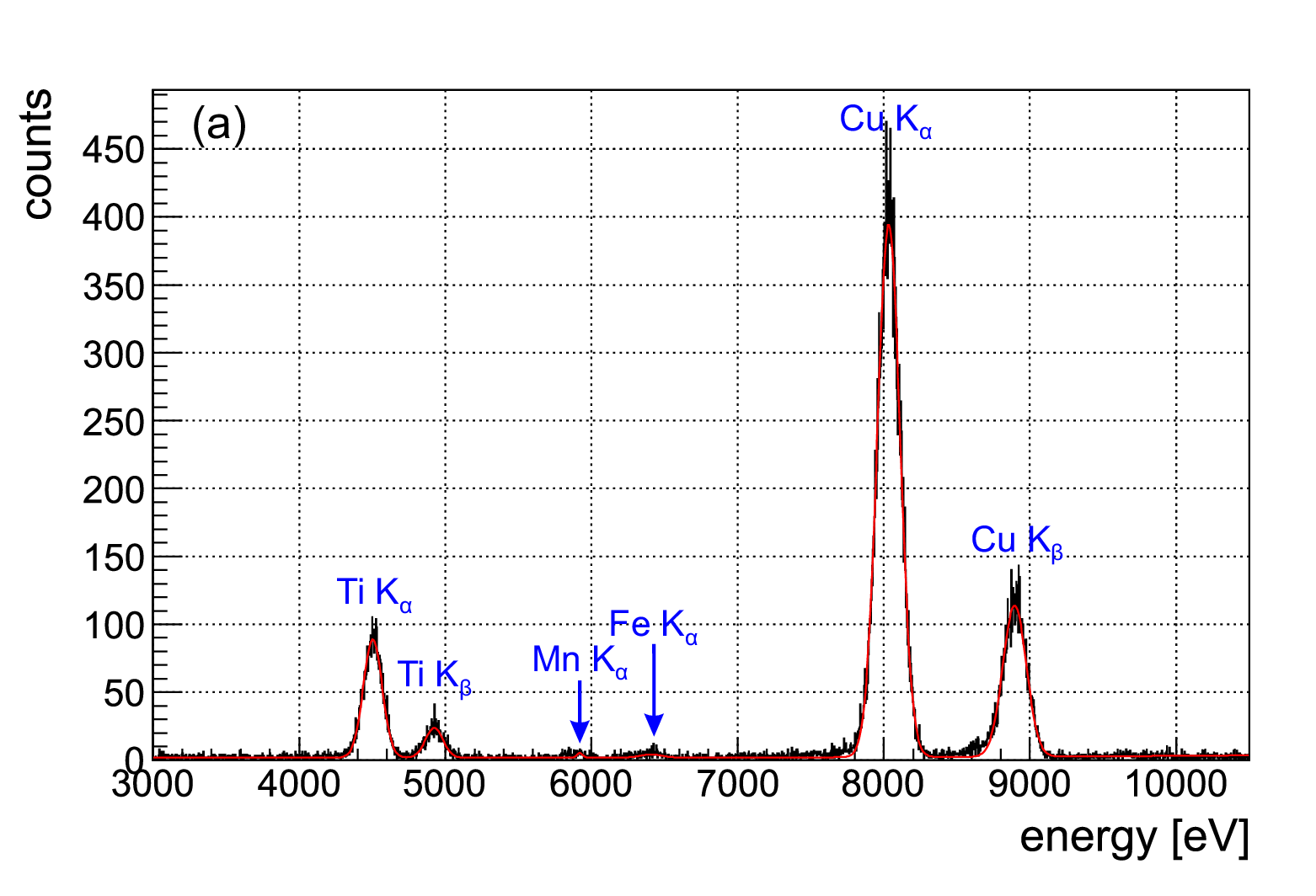}
\includegraphics[width=0.49\textwidth]{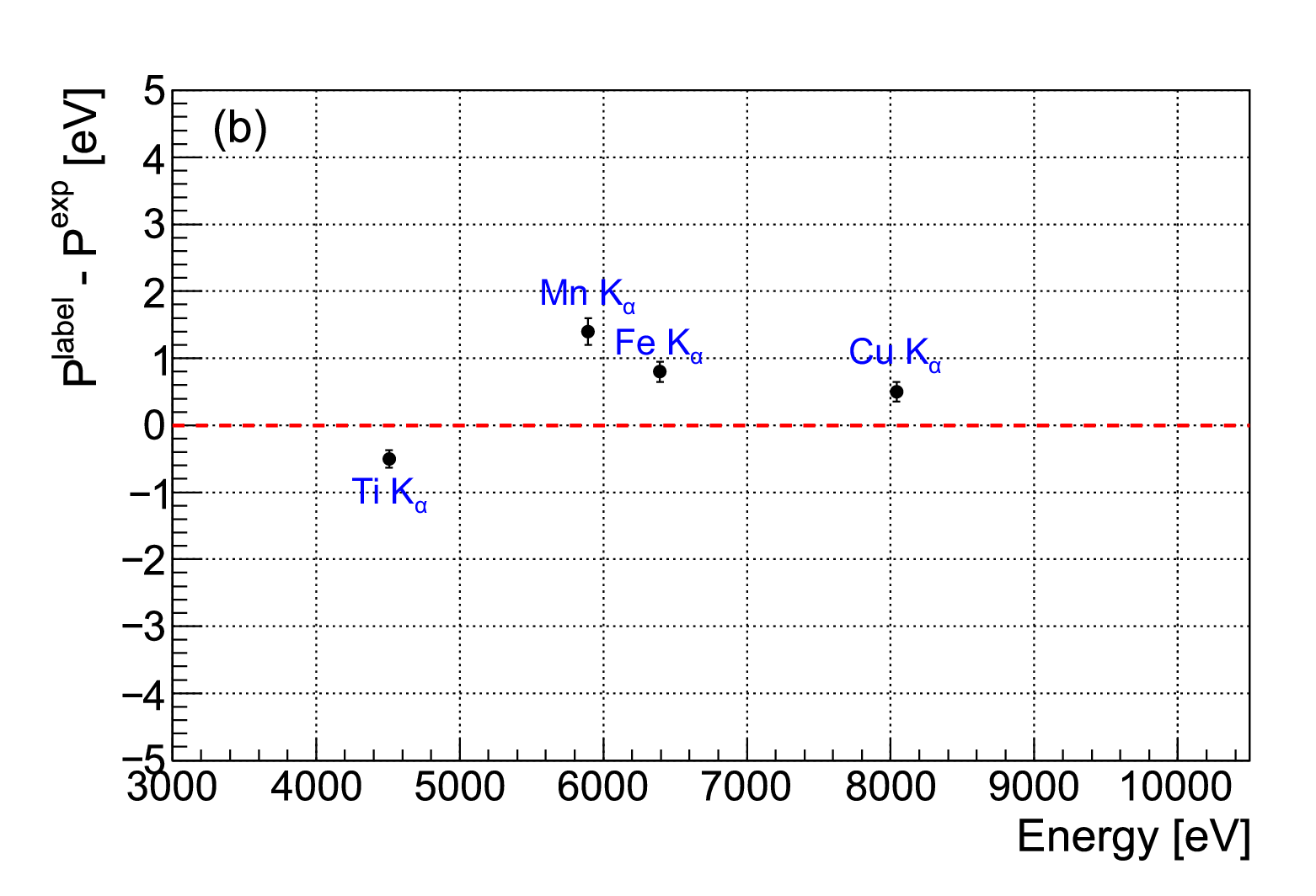}}
\vspace{-3pt}
\caption{(a) Calibrated energy spectrum of the single SDD. 
(b) Difference between each K$_{\alpha}$ experimental calibrated point ($P^{exp}_{i}$) with respect to its corresponding theoretical value ($P^{label}_{i}$).} 
\label{Fig:hRes}
\end{figure}

The final energy spectrum is shown in Fig.~\ref{Fig:hRes}(a).
The difference between each K$_{\alpha}$ experimental calibrated point ($P^{exp}_{i}$) with respect to its corresponding theoretical value ($P^{label}_{i}$) was used for the evaluation of the system linearity:
\begin{equation}
    R_i = P^{label}_{i} - P^{exp}_{i}.
\end{equation}

\noindent 
The distribution in Fig. \ref{Fig:hRes}(b) shows that the peak energies differ from the theoretical values by less than 1.5 eV, setting the linear response $\Delta E/E$ of the SDD below $10^{-3}$.

\section{Stability}

\begin{figure}[htb]
\centerline{%
\includegraphics[width=0.49\textwidth]{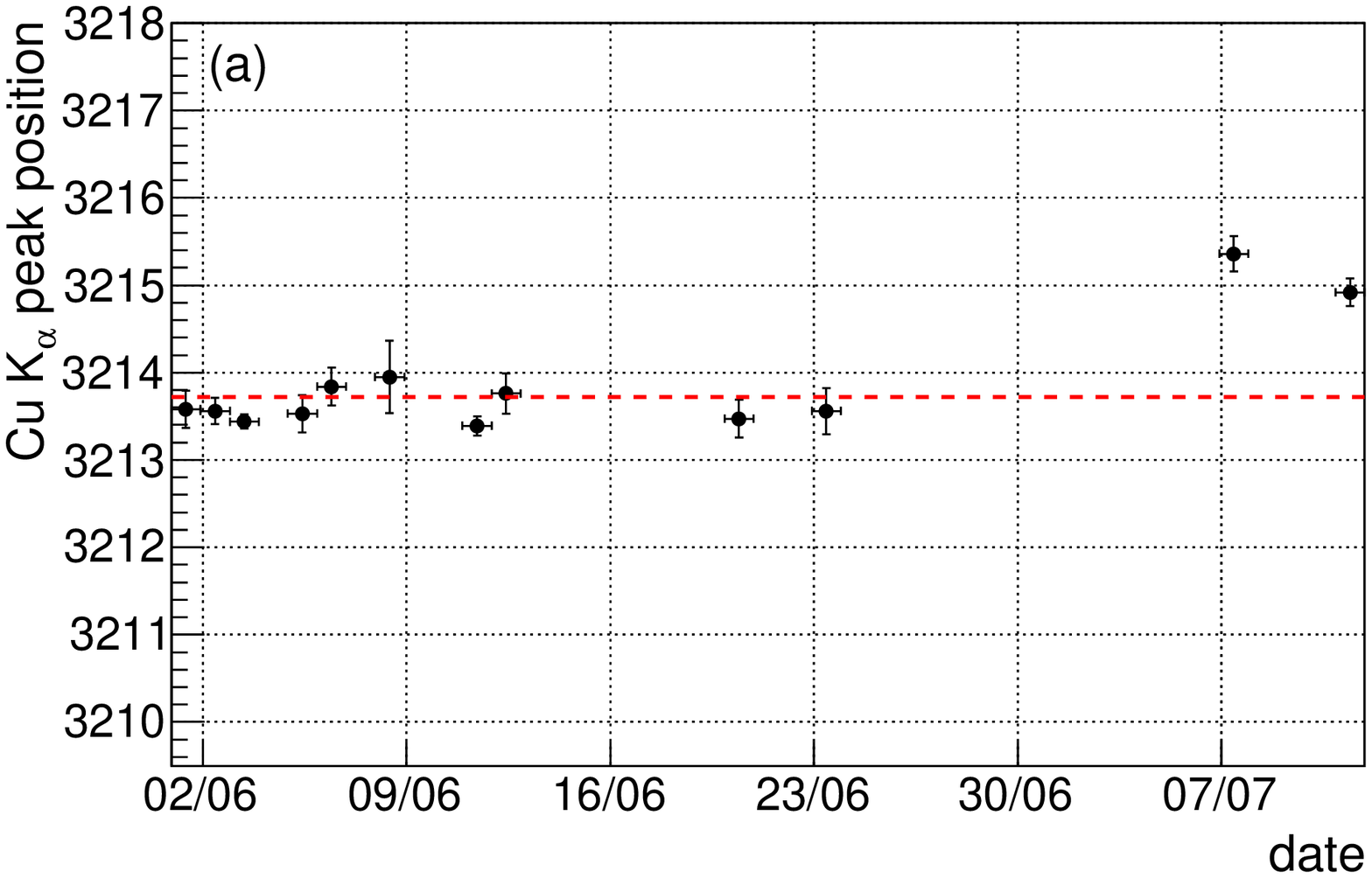}
\includegraphics[width=0.49\textwidth]{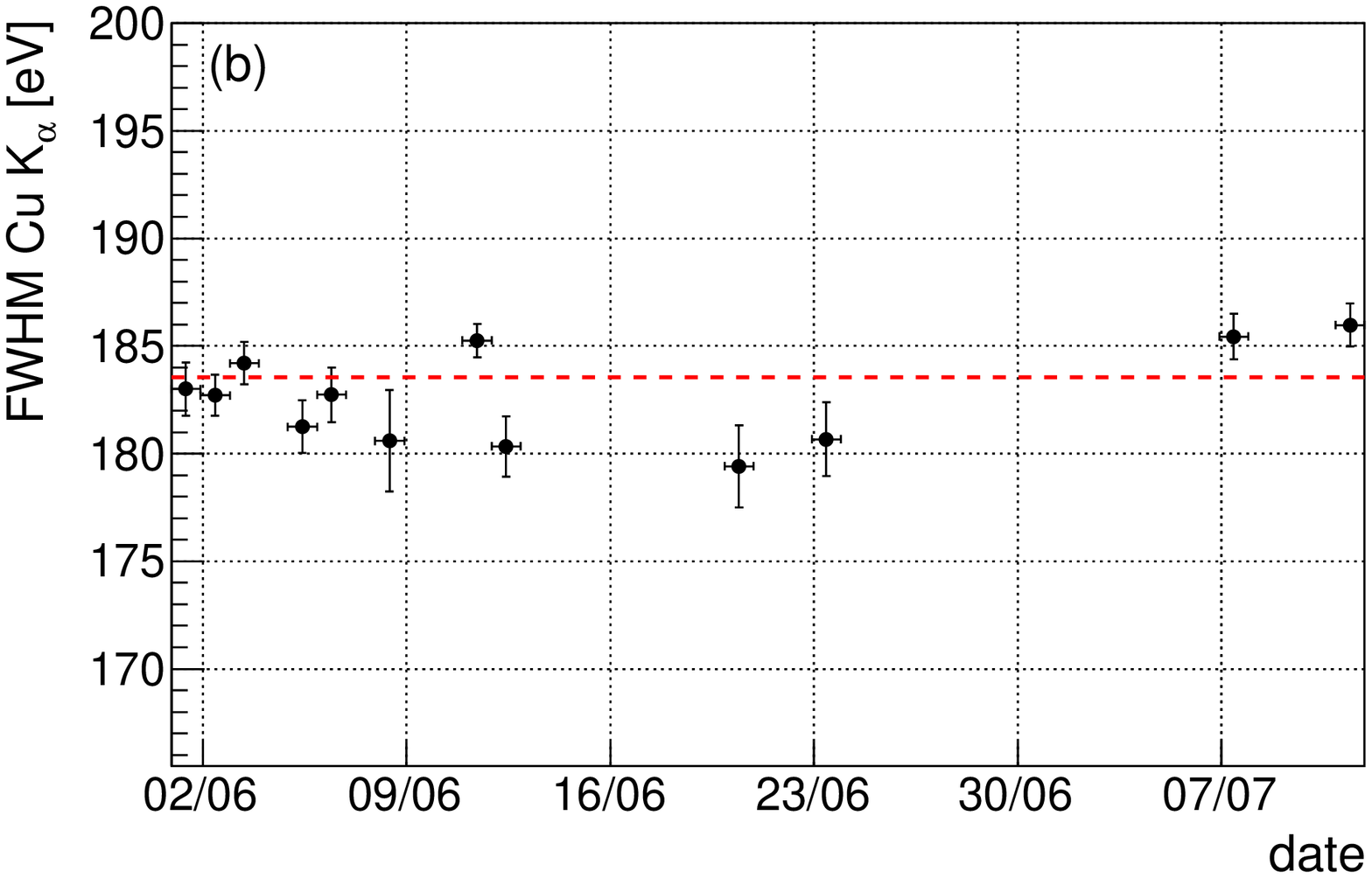}}
\caption{(a) The position of the Cu K$_{\alpha}$ peak (in arbitrary ADC units) as a function of time for a typical SDD. (b) Width stability of the  Cu K$_{\alpha}$ line.}
\label{Fig:hStab}
\end{figure}

Another important aspect of the SDD system that needs to be monitored during high-precision measurements is its long-term stability, since it is a possible source of systematic error in the experiment.
During the kaonic deuterium test measurement period, which took
place in June-July 2022, 12 calibration runs were performed. 
For each SDD, fluctuations over time of the Cu K$_{\alpha}$ peak position were used to control the stability (see Fig. \ref{Fig:hStab}(a)). 
%
During the measurements in June 2022 the fluctuations of the Cu K$_{\alpha}$ position were about 0.5 channels, which corresponds to $\sim$1.5~eV, and are comparable with the residuals evaluated during the analysis in the Section \ref{Sec:Linearity}.
However, the deviation of the peak position in the last two runs of July was larger (within \mbox{$\sim1.5$~channels}, corresponding to $\sim4.5$~eV), which is related to more unstable beams condition during the measurement.

The energy resolution reflects its accuracy in determining the energy of the incoming radiation. 
For the SDD, the Full Width at Half Maximum (FWHM) of the peak depends on the energy of the incoming photons, and on the electronic noise of the system~\cite{Miliucci_CM2019}.
Fig. \ref{Fig:hStab}(b) shows the FWHM (in~eV) of the Cu K$_{\alpha}$ line, as a function of time, for a typical SDD unit.
The width of the peak varies by about $\pm 2.5$~eV. 

Both results indicate very good stability of the SDDs throughout the experiment, thus providing an excellent opportunity to accurately determine the position of the kaonic deuterium peak, as well as the peak width.

\section*{Conclusions}
The SIDDHARTA-2 collaboration has developed a new Silicon Drift Detector (SDD) system for the measurement of the kaonic deuterium X-ray transitions to the ground state. 
The experimental setup has been installed at the DA$\Phi$NE collider of the INFN-LNF.  
The linearity and stability of the SDDs were monitored during the kaonic deuterium test run, and its performance is described in this work. 
The detectors are linear and stable within a few eV in the energy region of interest. 
Following the 2022 kaonic deuterium test run, a full measurement will be performed in 2023 to extract for the first time the strong interaction effects in this exotic atom. 


\section*{Acknowledgments}

We thank C. Capoccia from LNF-INFN and H. Schneider, L. Stohwasser, and D. Pristauz-Telsnigg from Stefan-Meyer-Institut for their fundamental contribution in designing and building the SIDDHARTA-2 setup. 
We thank as well the DA$\Phi$NE staff for the excellent working conditions and permanent support. 
Part of this work was supported by the Austrian Science
Fund (FWF): [P24756-N20 and P33037-N]; 
the Croatian Science Foundation under the project IP-2018-01-8570; 
the EU STRONG-2020 project (Grant Agreement No. 824093); 
the EU Horizon 2020 research and innovation programme under the Marie Sk{\l}odowska-Curie Grant Agreement No. 754496; 
the Japan Society for the Promotion of Science JSPS KAKENHI Grant No. JP18H05402;
the Polish National Agency for Academic Exchange (Grant No. PPN/BIT/2021/1/00037);
the SciMat and qLIFE Priority Research Areas budget under the program Excellence Initiative - Research University at the Jagiellonian University.

\printbibliography

\end{document}